\documentclass[preprint,12pt,number]{elsarticle}

\usepackage[T1]{fontenc}
\usepackage[utf8]{inputenc}
\usepackage{lmodern}
\usepackage{microtype}

\usepackage[a4paper,margin=2.5cm]{geometry}   
\usepackage{setspace}
\usepackage[mathlines]{lineno}

\usepackage{amsmath,amssymb}
\usepackage{bigints}

\usepackage{graphicx}
\usepackage{array}                 
\usepackage{booktabs}
\usepackage{longtable}
\usepackage[font=small,labelfont=bf]{caption}
\usepackage{xcolor}                
\usepackage{tikz}
\usetikzlibrary{positioning,arrows.meta,shapes.misc,fit,backgrounds}

\usepackage[colorlinks=true,citecolor=blue,linkcolor=blue,urlcolor=blue]{hyperref}
\usepackage{xurl}
\usepackage{orcidlink}            

\biboptions{sort&compress}
\definecolor{bluePolimi}{RGB}{22, 44, 80}
\definecolor{lightBluePolimi}{RGB}{91, 122, 172}
\definecolor{greenPolimi}{RGB}{0, 110, 0}
\definecolor{redPolimi}{RGB}{180, 0, 0}

\newcolumntype{M}[1]{>{\centering\arraybackslash}m{#1}}

\tikzset{
    beamColor/.style={red!80!black},
    neutronColor/.style={blue!60!cyan},
    targetBody/.style={top color=gray!10, bottom color=gray!40, middle color=white, shading angle=0},
    targetFace/.style={fill=gray!20, draw=gray!60},
    controlSurf/.style={dashed, color=gray!40, thin},
    myLabel/.style={font=\sffamily\small, text=black!80}
}

\journal{Annals of Nuclear Energy}

\begin{document}

\begin{frontmatter}

\title{Simulating scintillation detector responses with OpenMC and pyMPPost: a track-based coupling verified against MCNPX-PoliMi}

\author[polimi,umich]{Lorenzo Loi\texorpdfstring{\,\orcidlink{0009-0003-7904-3731}}{}}
\author[umich]{David Breitenmoser\texorpdfstring{\,\orcidlink{0000-0003-0339-6592}}{}}
\author[umich]{Shaun D. Clarke\texorpdfstring{\,\orcidlink{0000-0003-2469-3166}}{}}
\author[umich]{Sara A. Pozzi\texorpdfstring{\,\orcidlink{0000-0001-6827-3652}}{}}
\author[polimi,ku,entc]{Antonio~Cammi\texorpdfstring{\,\orcidlink{0000-0003-1508-5935}\corref{cor1}}{}}
\cortext[cor1]{Corresponding author}
\ead{antonio.cammi@polimi.it}

\affiliation[polimi]{organization={Politecnico di Milano, Department of Energy},
            addressline={Via La Masa 34},
            city={Milano},
            postcode={20133},
            country={Italy}}
\affiliation[umich]{organization={Department of Nuclear Engineering and Radiological Sciences, University of Michigan},
            city={Ann Arbor},
            postcode={48109-2104},
            state={MI},
            country={USA}}
\affiliation[ku]{organization={Department of Mechanical and Nuclear Engineering, Khalifa University},
            city={Abu Dhabi},
            postcode={127788},
            country={United Arab Emirates}}
\affiliation[entc]{organization={Emirates Nuclear Technology Centre (ENTC), Khalifa University of Science and Technology},
            city={Abu Dhabi},
            postcode={127788},
            country={United Arab Emirates}}

\begin{abstract}
Simulations of the detector response are essential for the design and analysis of neutron and photon measurements in nuclear nonproliferation, safeguards, and active interrogation studies. MCNPX-PoliMi, combined with the post-processor MPPost and its Python implementation pyMPPost, is a well-established and validated framework for this purpose. Its detector models, however, are not available to users of the open-source Monte Carlo code OpenMC, which does not write an equivalent collision-level output. This work couples OpenMC to pyMPPost, allowing the open-source community to use the detector models already implemented and validated in pyMPPost without modifying either code. A Python class, \texttt{TrackAnalyzer}, reads the particle tracks written by OpenMC, reconstructs the interactions in the detector material along with the target nuclide and writes them in the MCNPX-PoliMi collision file format. The method relies only on the track output and is therefore applicable to any OpenMC version that writes particle tracks, including the development branch with photonuclear physics used in this work. The coupling was verified code to code against MCNPX-PoliMi for a trans-stilbene and a $^{4}$He scintillator exposed to $^{252}$Cf and deuterium-deuterium fusion neutron sources, and for a NaI(Tl) scintillator exposed to $^{137}$Cs and $^{152}$Eu photon sources. The total number of pulses agrees within $\pm$1\% in all cases, and the light-output spectra agree within a few percent in the covered spectral domain. By making the detector response models of pyMPPost available within a fully open-source simulation chain, this work extends detector response studies with OpenMC to nuclear nonproliferation and active interrogation applications, including those involving photonuclear reactions, and provides an accessible tool for education and training in radiation detection.
\end{abstract}




\end{frontmatter}


\section{Introduction}
\label{sec:introduction}

Neutron and photon detectors are central to nuclear non-proliferation and safeguards, where they are used to detect, characterize and verify special nuclear material in both passive measurements and active interrogation~\cite{IAEA_STR385,RUNKLE201275,POZZI2012119,Clarke2024}. Organic scintillators such as trans-stilbene are widely used for fast-neutron and gamma-ray detection in these applications~\cite{SHIN201936}, high-pressure $^{4}$He scintillators have been applied to the detection of photofission neutrons~\cite{Searfus_PRA}, and inorganic scintillators such as NaI(Tl) remain a standard technology for gamma-ray spectroscopy~\cite{DIFULVIO2016169}. The design of new detection systems and the interpretation of measurements require Monte Carlo simulations that predict the detector response itself, including the conversion of deposited energy into light, the energy and time resolution of the detector, and the formation of pulses, rather than only the particle flux or the energy deposited in the detector volume. The same need arises more broadly in nuclear engineering, wherever neutron and gamma-ray detectors are designed or their measurements interpreted, for example for radiation monitoring at research reactors, accelerators and fusion facilities.

For this purpose, the MCNP-PoliMi code~\cite{POZZI2003550} and its successor MCNPX-PoliMi~\cite{POZZI2012119,PoliMiManual2012} were developed to simulate each particle interaction as closely as possible to the physical process. During transport, MCNPX-PoliMi writes a collision file that lists, for every history, each interaction occurring in the detector cells together with the reaction type, the target nucleus, the deposited energy, the time and the position. The post-processor MPPost~\cite{Miller2012,MPPostManual2012} uses this information to compute the response of organic and inorganic scintillators and $^{3}$He detectors, including light output with ionization quenching, energy and time resolution and pulse generation, and to reproduce common analysis techniques such as time-of-flight, cross-correlation and multiplicity counting. These capabilities are now available in pyMPPost~\cite{pyMPPost}, its Python implementation. The framework is supported by measurements: MCNPX-PoliMi with MPPost reproduced measured pulse-height distribution of a $^{252}$Cf source in an EJ-309 liquid scintillator~\cite{Miller2012}, the response to mixed-oxide fuel and plutonium-oxide samples~\cite{POZZI2012119}, and the total neutron count rate measured with a trans-stilbene detector within $\pm$3\%~\cite{SHIN201936}. Other tools address the same problem, for example the DRiFT toolkit, which processes the output of MCNP~\cite{Andrews2018DRiFT,Kulesza2022}, and the Geant4 toolkit~\cite{AGOSTINELLI2003250}. MCNPX-PoliMi and MPPost, however, remain a well-established reference for detector response simulations in nonproliferation and safeguards.

OpenMC~\cite{ROMANO201590} is an open-source Monte Carlo particle transport code with continuous-energy neutron and photon transport~\cite{LUND2018} and a Python interface for building models and analyzing results. Being freely available and easy to modify, it is well suited to teaching and to research, and it is increasingly used in reactor physics, shielding and fusion neutronics, also as an independent code for the verification of established ones. It can therefore broaden the range of analyses that can be performed when used alongside the reference tools. MCNPX-PoliMi, instead, is distributed by RSICC under license, together with MCNP~\cite{PoliMiManual2012}, which limits its availability in some settings. In addition, a development branch of OpenMC includes photonuclear physics~\cite{OpenMCphotonuclear}, which is of direct interest for active interrogation with photon beams~\cite{RUNKLE201275}. Validating these capabilities against measurements requires the simulation of the detector response, and therefore a way to obtain it with OpenMC.

OpenMC does not write by default an MCNPX-PoliMi collision file. It can, however, write the complete sequence of states of each particle history, and these track files can be read through the Python API from version 0.13.1 onward~\cite{OpenMC_0131}. The \textit{collision tracking} feature introduced in the official release 0.15.3 records the reaction type and the target nuclide of each collision~\cite{Saliba2026}, but it is not available in the development branch with photonuclear physics. A significant effort has been devoted to the development of MPPost and pyMPPost, and the detector models they implement have been tested extensively against measurements~\cite{POZZI2012119, SHIN201936, Miller2012, breitenmoser2026disentanglingmixedneutronfields}. Rather than reimplementing these models for a new transport code, it is therefore convenient to couple OpenMC to the existing post-processor by producing its input file.

The present work is focused on this coupling. A Python class, \texttt{TrackAnalyzer}, reads the OpenMC track file, reconstructs the interactions in the detector material and writes them in the MCNPX-PoliMi collision file format, which pyMPPost reads without any modification. Since the track file does not store the target nuclide, the nuclide of each neutron collision is reconstructed from two-body elastic kinematics, which is required for detectors containing several nuclides with a nonlinear light response. The method relies only on the track output and is applicable to any OpenMC version that writes particle tracks ($\geq$ 0.13.1), including the current stable release (0.15.3) and the development branch with photonuclear physics used for the simulations of this work. The coupling is verified code to code against MCNPX-PoliMi for both neutrons and photons: a trans-stilbene scintillator, with several nuclides and a nonlinear response, and a $^{4}$He scintillator, with a single nuclide and a linear response, exposed to $^{252}$Cf and DD neutron sources; and a NaI(Tl) scintillator exposed to $^{137}$Cs and $^{152}$Eu photon sources. In this way, detectors of standard technology can also be analyzed within the same open-source environment. Since the two simulation chains use the same nuclear data and the same detector models in pyMPPost, the comparison of the detector responses also provides an indirect verification of OpenMC against MCNPX-PoliMi for the quantities that determine them: the type, position, time and deposited energy of the collisions in the detector. The reconstruction is further checked event by event against the collision tracking feature of the official release. \texttt{TrackAnalyzer} is released as open-source software~\cite{TrackAnalyzerRepo}, so that the coupling can be used, extended and verified by the OpenMC community.

The paper is organized as follows. Section~\ref{sec:methodology} describes the reconstruction of the interactions from the track file and the simulation settings. Section~\ref{sec:results} compares the detector responses obtained with the two codes, and Section~\ref{sec:discussion} discusses the residual differences, the limitations of the method and its computational cost. Conclusions are drawn in Section~\ref{sec:conclusion}. ~\ref{app:collision-track} reports the event-by-event comparison with the collision tracking feature.

\section{Methodology}
\label{sec:methodology}

\begin{figure*}[!t]
\centering
\resizebox{\textwidth}{!}{
\begin{tikzpicture}[
  font=\sffamily,
  >={Stealth[length=2.2mm,width=1.6mm]},
  box/.style={draw=black!65, fill=white, line width=0.6pt, rounded corners=2pt,
              minimum width=26mm, minimum height=14mm, text width=24mm,
              align=center, inner sep=1.5pt},
  wide/.style={minimum width=30mm, text width=28mm},
  doc/.style={chamfered rectangle, chamfered rectangle corners=north east,
              chamfered rectangle xsep=2mm, draw=black!65, fill=black!4,
              line width=0.6pt, minimum width=24mm, minimum height=14mm,
              text width=22mm, align=center, inner sep=1.5pt},
  work/.style={draw=cWork, fill=cWork!8, line width=0.9pt, rounded corners=2pt,
               minimum width=38mm, minimum height=24mm, text width=35mm,
               align=left, inner sep=2pt},
  flow/.style={->, line width=0.7pt, draw=black!75},
  note/.style={font=\sffamily\scriptsize\itshape, text=black!65, align=center},
]
\definecolor{cWork}{HTML}{D55E00}
\providecommand{\ttl}[1]{{\small\bfseries #1}}
\providecommand{\sub}[1]{{\scriptsize\color{black!65}#1}}

\node[box, wide] (openmc) at (-2mm,0)
  {\ttl{OpenMC}\\[1pt]\sub{transport with}\\[-1.5pt]\sub{track output}};
\node[doc] (tracks) at (31mm,0)
  {\ttl{Track file}\\[1pt]\sub{\ttfamily tracks.h5}};
\node[work] (coupler) at (67mm,0)
  {{\centering\ttl{TrackAnalyzer}\par}\vspace{2pt}
   \sub{1.\ collision detection}\\[-1.5pt]
   \sub{2a.\ target nuclide (n)}\\[-1.5pt]
   \sub{2b.\ interaction type ($\gamma$)}\\[-1.5pt]
   \sub{3.\ deposited energy}\\[-1.5pt]
   \sub{4.\ PoliMi-format writer}};

\node[box, wide] (polimi) at (-2mm,-27mm)
  {\ttl{MCNPX-PoliMi}\\[1pt]\sub{reference}\\[-1.5pt]\sub{transport code}};

\node[doc, minimum height=41mm] (coll) at (104mm,-13.5mm)
  {\ttl{Collision file}\\[1pt]\sub{MCNPX-PoliMi}\\[-1.5pt]\sub{format}};
\node[box] (pymppost) at (134mm,-13.5mm)
  {\ttl{pyMPPost}\\[1pt]\sub{detector response}\\[-1.5pt]\sub{post-processor}};
\node[doc] (resp) at (164mm,-13.5mm)
  {\ttl{Light output}\\[1pt]\sub{distributions}};

\draw[flow] (openmc) -- (tracks);
\draw[flow] (tracks) -- (coupler);
\draw[flow] (coupler.east) -- (coupler.east -| coll.west);
\draw[flow] (polimi.east) -- node[note, above]{collision file written during transport} (polimi.east -| coll.west);
\draw[flow] (coll) -- (pymppost);
\draw[flow] (pymppost) -- (resp);
\end{tikzpicture}}
\caption{Workflow for computing the detector response with OpenMC and pyMPPost. \texttt{TrackAnalyzer}, developed in this work, reconstructs the interactions in the detector from the OpenMC track file and writes them in the MCNPX-PoliMi collision file format, which MCNPX-PoliMi produces natively during transport. The collision files of both codes are processed by pyMPPost with identical parameters.}
\label{fig:workflow}
\end{figure*}
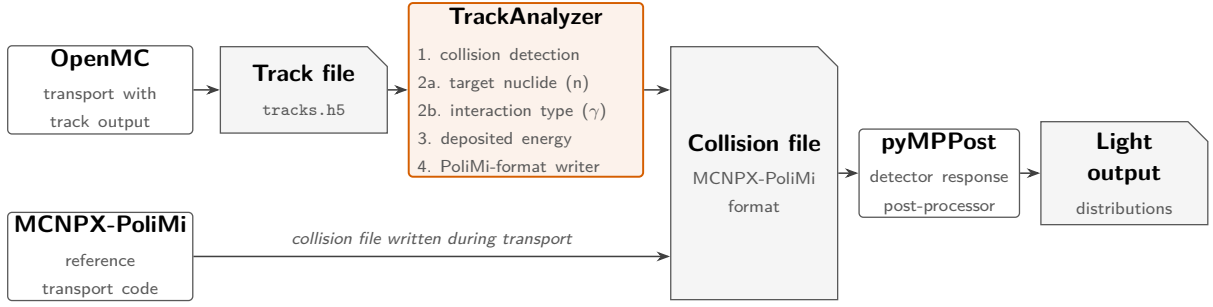

The complete simulation chain is illustrated in Fig.~\ref{fig:workflow}. \texttt{TrackAnalyzer} reads the OpenMC track file, reconstructs the interactions occurring in the detector material (Sections~\ref{sec:tracks}--\ref{sec:photon-reco}) and writes them in the MCNPX-PoliMi collision file format (Section~\ref{sec:collision-file}); the resulting file is then processed by pyMPPost in exactly the same manner as a file generated by MCNPX-PoliMi (Section~\ref{sec:pympost}). Throughout this work, MCNPX-PoliMi serves as the reference against which the coupling is verified.

All OpenMC simulations presented in this work were performed with a development branch of OpenMC (version 0.16.1-dev138) that includes photonuclear physics~\cite{OpenMCphotonuclear}; this version was adopted in view of future validation work on photonuclear reactions. Since the reconstruction relies solely on the track output, accessed through the \texttt{openmc.Track} class, and requires no modification of the transport code, it is applicable to any OpenMC version from 0.13.1 onward, including the current stable release (0.15.3).

\subsection{Particle tracks and collision detection}
\label{sec:tracks}

When track output is enabled, OpenMC writes for every source particle and for all its secondary particles a sequence of states, each containing the position $\mathbf{r}$, the direction $\mathbf{u}$, the kinetic energy $E$, the time $t$, the statistical weight $w$, and the cell and material in which the particle is located. A new state is recorded after each event: birth, surface crossing, collision and death. The state therefore describes the particle immediately after the event.

The track states do not specify the type of event that generated them, nor, for a collision, the reaction type or the target nuclide. For photons this information is not needed to compute the light output, since the energy is deposited by electrons regardless of the target atom. For neutrons in organic scintillators, instead, the light output depends strongly on the recoiling nucleus: because of ionization quenching, described by Birks' law (Section~\ref{sec:pympost}), a proton recoil produces much more light than a carbon recoil of the same energy~\cite{Knoll:1300754}. Writing a collision file for pyMPPost therefore requires the target nuclide of each neutron collision to be reconstructed indirectly from the available states. The method developed for this purpose is described in Section~\ref{sec:neutron-reco}. The collision tracking feature of the official release~\cite{Saliba2026}, which records this information directly, is used in~\ref{app:collision-track} as a reference to verify the reconstruction event by event.

A collision is identified as a change of direction between two consecutive states $i$ and $i+1$, i.e.\ when the components of $\mathbf{u}_i$ and $\mathbf{u}_{i+1}$ differ by more than a numerical tolerance of about $10^{-5}$. Surface crossings do not change the direction and are ignored. Since each state is written after the corresponding event, the collision takes place at state $i+1$: the position, time, cell and material of the interaction are taken from state $i+1$, while the incident energy and the weight of the particle are taken from state $i$. Only collisions occurring in the detector material are retained. A direction change caused by a reflective boundary would be indistinguishable from a collision, so reflective surfaces in contact with the detector material must be avoided. Since the detector response is normalized per source particle, the track file must contain all the simulated histories, which in OpenMC requires the maximum number of written tracks (\texttt{max\_tracks}) to be at least equal to the number of histories.

\subsection{Neutron interactions}
\label{sec:neutron-reco}

The target nuclide is reconstructed using only the energies and directions stored in the track states. The method is based on the kinematics of two-body elastic scattering, which relates the energy lost by the neutron to its scattering angle and to the mass of the target. For a neutron scattered elastically by a free nucleus at rest with mass ratio $A$ (atomic weight ratio to the neutron mass), the ratio between the outgoing and incident neutron energies at a laboratory scattering angle $\theta$ is~\cite{Duderstadt1976}
\begin{equation}
\frac{E'}{E} = \left[\frac{\cos\theta + \sqrt{A^{2} - \sin^{2}\theta}}{A+1}\right]^{2}.
\label{eq:elastic}
\end{equation}
Eq.~\eqref{eq:elastic} holds under three assumptions. The kinematics is non-relativistic, which is adequate for the neutron energies considered here (below 20~MeV). The collision is elastic, with no energy spent in exciting the target. The target is free and at rest, so that its thermal motion and chemical binding are neglected; this requires the neutron energy to be much larger than the thermal energy of the target, i.e.\ above a few eV. 

The method requires the list of candidate nuclides to be defined before the analysis, on the basis of the composition of the detector material. Each candidate is specified by its atomic weight ratio $A$, taken from the nuclear data used in the transport simulation, and by the ZAID to be written in the collision file. The list must include every nuclide on which a significant number of collisions can occur, since a collision on a nuclide that is not listed cannot be identified correctly. For each collision, $\cos\theta = \mathbf{u}_i\cdot\mathbf{u}_{i+1}$ and $E'/E = E_{i+1}/E_i$ are obtained from the track states. Before the kinematic test, three classes of collisions are rejected: collisions with an incident energy below 10~eV, where the assumption of a target at rest no longer holds; collisions with $E'/E > 1 + 10^{-5}$, i.e.\ thermal up-scattering; and collisions with $E'/E > 1 - 10^{-4}$, whose energy loss is too small for the candidates to be discriminated and whose contribution to the deposited energy is negligible. For the remaining collisions, Eq.~\eqref{eq:elastic} is evaluated for every candidate. A preliminary check discards the candidate if the measured ratio is below the minimum reachable value $\alpha = [(A-1)/(A+1)]^{2}$, or, for candidates with $A\leq1$ such as $^{1}$H, if the scattering angle exceeds the maximum allowed by elastic kinematics. Among the remaining candidates, the one with the smallest residual, defined as the absolute difference between the predicted and the measured energy ratio, is selected. If this residual does not exceed a tolerance of 0.05, the collision is written as an elastic scattering event (PoliMi interaction code $-99$) with the ZAID of that nuclide and a deposited energy $E_i - E_{i+1}$; otherwise, as when no candidate passes the preliminary check, the collision is considered not compatible with elastic scattering and is not written. The same thresholds and tolerance were used for all the neutron cases. Collisions that pass the kinematic test are then subject to the threshold on the deposited energy described in Section~\ref{sec:collision-file}.

The discrimination between candidates is most effective when their masses are very different, as for hydrogen and carbon in organic scintillators, because the predicted energy ratios then differ markedly at any scattering angle. Candidates with similar masses would be harder to separate, but in scintillators this has little practical effect, since heavy recoils produce little light. For a detector made of a single nuclide, such as $^{4}$He, the identification is trivial, but the same kinematic test is used to reject collisions that are not compatible with elastic scattering.

Inelastic scattering and other non-elastic reactions are not written. For the detectors considered in this work, this is a minor limitation: in trans-stilbene, neutron inelastic scattering on carbon is possible only above about 4.8~MeV, and in $^{4}$He no non-elastic channel is open below 20~MeV. The accuracy of the identification is quantified event by event in~\ref{app:collision-track}.

\subsection{Photon interactions}
\label{sec:photon-reco}

Photon transport was performed with the thick-target bremsstrahlung (TTB) approximation, in which electrons and positrons are not transported: their energy is deposited locally, and only the secondary photons they produce (bremsstrahlung, fluorescence and annihilation photons) are tracked and appear in the track file. Photon interactions are classified from the change of energy and direction between consecutive states and from the secondary photons born at the interaction site. A direction change without energy change is a coherent scattering (PoliMi code 2), which deposits no energy and is not written. A direction change with energy loss is a Compton scattering (code 1). A track whose last state has zero energy corresponds to an absorption: if the incident energy exceeds $2m_ec^{2} = 1.022$~MeV and two photons with energy within 0.1\% of 511~keV are born at the end point, the event is a pair production (code 5); otherwise it is a photoelectric absorption (code 3). A track ending with non-zero energy corresponds to a photon leaving the detector or reaching the energy cutoff, and nothing is written. Code 4 of MCNPX-PoliMi (capture with double fluorescence) cannot be distinguished from code 3 using track data and is never written.

The energy deposited at an interaction site is
\begin{equation}
E_{\mathrm{dep}} = E_{\mathrm{in}} - E_{\mathrm{out}} - \sum_{k} E_{k},
\label{eq:edep}
\end{equation}
where $E_{\mathrm{in}}$ and $E_{\mathrm{out}}$ are the photon energies before and after the interaction ($E_{\mathrm{out}} = 0$ for absorptions) and $E_{k}$ are the energies of the secondary photons born at the same position, identified within a tolerance of $10^{-6}$~cm. The secondary photons are tracked independently and can produce their own interactions in the detector, so their energy must be subtracted to avoid double counting. Interactions with incident energy below 1~keV are not written.

The energy balance was checked on the $^{137}$Cs case: for histories in which no photon escapes the crystal, the sum of the deposited energies equals the source energy, 661.7~keV. Pair production is implemented but was not specifically verified: although about 22\% of the photons of the $^{152}$Eu source have energies above 1.022~MeV, at these energies the pair production cross section of NaI is small compared with that of Compton scattering, so that few pair production events occur.

\subsection{Collision file}
\label{sec:collision-file}

Each reconstructed interaction is written as one line of the MCNPX-PoliMi collision file, with the fixed-width layout described in the MCNPX-PoliMi user's manual~\cite{PoliMiManual2012} and read by pyMPPost (see Table~\ref{tab:collision-file}). Some columns are filled differently from MCNPX-PoliMi. The particle number, which in MCNPX-PoliMi distinguishes the particles produced in the same source event, is the index of the particle track within the OpenMC history. The number of scatterings counts the events written for the particle in the detector, whereas MCNPX-PoliMi counts the elastic scatterings (neutrons) or Compton scatterings (photons) undergone by the particle since its birth. The code column, which in MCNPX-PoliMi encodes the inelastic and $(n,xn)$ history of neutrons and the reaction number of photons, is not reconstructed and is set to zero. The generation number is also set to zero, which coincides with the MCNPX-PoliMi value for the cases considered here, since no fission occurs in the detectors. For photons, the target nucleus cannot be identified from the tracks and a single ZAID is written for the detector material. None of these columns affects the light output: pyMPPost computes the light of each pulse from the particle type, the interaction type, the target nucleus, the deposited energy and the time of the collisions in each detector cell~\cite{Miller2012}. 
Two options of the MCNPX-PoliMi input are reproduced when the file is written, so that the two collision files contain the same kind of events. Collisions that deposit less than a given energy, specified separately for neutrons and photons as with the RPOL card, are not written, and the output can be restricted to a list of detector cells, as with IPOL(7)--(8)~\cite{PoliMiManual2012}. This threshold on the deposited energy is distinct from the threshold on the incident energy of Section~\ref{sec:neutron-reco}: the former removes collisions that release little energy, the latter collisions of neutrons too slow for Eq.~\eqref{eq:elastic} to apply.

\begin{table}[!t]
\centering
\small
\caption{Columns of the collision file written by \texttt{TrackAnalyzer}, following the MCNPX-PoliMi format \cite{PoliMiManual2012}.}
\label{tab:collision-file}
\begin{tabular*}{\columnwidth}{@{\extracolsep{\fill}}rl@{}}
\toprule
Column & Content \\
\midrule
1 & History number \\
2 & Particle number \\
3 & Particle type (1 neutron, 2 photon) \\
4 & Interaction type ($-99$; 1, 3, 5) \\
5 & Target ZAID \\
6 & Cell number \\
7 & Deposited energy (MeV) \\
8 & Time (shake) \\
9--11 & Position $x$, $y$, $z$ (cm) \\
12 & Generation number (0) \\
13 & Weight \\
14 & Number of scatterings \\
15 & Code (0) \\
16 & Incident energy (MeV) \\
\bottomrule
\end{tabular*}
\end{table}

\subsection{Detector response with pyMPPost}
\label{sec:pympost}

pyMPPost implements the algorithm of MPPost~\cite{pyMPPost, Miller2012}. For each history, it converts the energy deposited in every collision into scintillation light, sums the light of the collisions that occur in the same detector cell within a pulse generation time into a single pulse, and applies the energy and time resolution of the detector.

Organic scintillators exhibit a non-linear relationship between deposited energy and light output, which depends on the type of recoiling particle~\cite{Knoll:1300754,SHIN201936}. For proton recoils, the light output is described by a semi-empirical function based on Birks' formula~\cite{Birks1964},
\begin{equation}
L(E) = \int_{0}^{E}  \frac{S_{0}}{1 + k_{B}\,\dfrac{\mathrm{d}E}{\mathrm{d}x}}\,\mathrm{d}E,
\label{eq:birks}
\end{equation}
where $E$ is the initial recoil energy, $\mathrm{d}E/\mathrm{d}x$ is the proton stopping power in the scintillator, provided to pyMPPost as a table, and $S_0$ and $k_B$ are parameters fitted to measured light-output data. The light produced by carbon recoils is taken as 2\% of the deposited energy~\cite{BATCHELOR196170}. For electrons, pyMPPost assumes a light output proportional to the deposited energy, with unit proportionality constant, so that the light output is expressed in MeV electron equivalent (MeVee), consistently with the calibration of the detectors with monoenergetic gamma-ray sources~\cite{SHIN201936}. 

The energy resolution is applied as a Gaussian broadening of each pulse, with relative width~\cite{DIETZE1982549}
\begin{equation}
\frac{\Delta L}{L} = \sqrt{\delta_{1}^{2} + \frac{\delta_{2}^{2}}{L} + \frac{\delta_{3}^{2}}{L^{2}}},
\label{eq:resolution}
\end{equation}
where $L$ is the light output in MeVee and $\Delta L$ is the full width at half maximum (FWHM). The parameters $\delta_1$, $\delta_2$ and $\delta_3$ are usually determined by matching simulated and measured spectra of monoenergetic gamma-ray sources~\cite{SHIN201936}. In all cases the pulse generation time was 10~ns\footnote{The pulse generation time depends on the detector and on the acquisition system (e.g. the scintillation decay time and the integration gate), and should be set accordingly when simulations are compared with measurements. Since the aim here is a code-to-code verification, in which both transport codes are post-processed with identical parameters, the same value was used for all cases.}, the dead time zero, the time resolution 0.5~ns (standard deviation) and the lower light-output threshold 0.01~MeVee; the light-output distributions are histogrammed from 0.05~MeVee (Table~\ref{tab:cases}). Energy and time resolution were applied in all the simulations presented here. The detector-specific parameters are listed in Table~\ref{tab:detectors}.

\subsection{Verification cases}
\label{sec:cases}

The workflow was verified against MCNPX-PoliMi with pyMPPost on six cases, combining three detectors and four sources (Table~\ref{tab:cases}). No experimental data are used: the aim is a code-to-code verification of the coupling, in which the collision files of both codes are processed by the same version of pyMPPost with identical input parameters. The detectors were chosen so that each one tests a different part of the reconstruction:
\begin{itemize}
\item \textbf{Trans-stilbene}, an organic scintillator widely used for fast-neutron detection in nuclear safeguards and nonproliferation. It contains hydrogen and carbon, whose recoils produce very different amounts of light, and therefore tests the identification of the target nuclide described in Section~\ref{sec:neutron-reco}. The detector reproduces the $\varnothing 5.08 \times 5.08$~cm crystal characterized by Ref.~\cite{SHIN201936}, with density 1.160~g\,cm$^{-3}$ and composition 46.15~at\% H and 53.85~at\% C; the reflective wrapping, the aluminium housing and the photomultiplier tube are not modelled. The light output of proton recoils and the energy resolution are those obtained by Ref.~\cite{SHIN201936} with pulse integral analysis (150~ns integration window). The parameters of Eq.~\eqref{eq:birks}, $S_0 = 2.027$ and $k_B = 27.83$~mg\,cm$^{-2}$\,MeV$^{-1}$, were fitted to the light output measured for neutron energies between 0.79 and 4.57~MeV; above this range the function is extrapolated, which does not affect the comparison since the same function is used with both codes. The resolution parameters were obtained by matching simulated and measured $^{137}$Cs spectra.

\item \textbf{$\mathbf{^{4}}$He}, a high-pressure gas scintillator for fast-neutron detection and spectroscopy. It contains a single nuclide, so that no discrimination between candidates is needed, and it tests the reconstruction of elastic scattering on a target of intermediate mass, which can transfer up to 64\% of the neutron energy in a single collision. The detector is the S670 model, with an active volume 20~cm long and 4.95~cm in diameter, filled with $^{4}$He at 190~bar~\cite{SEARFUS2024169608, Searfus_PRA} and modelled with a density of 0.0269~g\,cm$^{-3}$. Its light output is proportional to the deposited energy, so that Eq.~\eqref{eq:birks} reduces to $L = E$ ($S_0 = 1$, $k_B = 0$) and 1~MeV of deposited energy corresponds to 1~MeVee. The resolution parameters were obtained by Ref.~\cite{SEARFUS2024169608} from a fit of the resolution in deposited energy between 0.0925 and 8.87~MeV.

\item \textbf{NaI(Tl)}, the most common inorganic scintillator for gamma-ray spectrometry. It tests the reconstruction of photon interactions described in Section~\ref{sec:photon-reco}: the high atomic number of iodine makes photoelectric absorption and the emission of K fluorescence X-rays frequent, which tests the subtraction of the energy of secondary photons in Eq.~\eqref{eq:edep}. The detector is a bare $3''\times3''$ crystal (diameter and height 7.62~cm), with density 3.67~g\,cm$^{-3}$ and a thallium concentration of 0.1~mol\%, corresponding to mass fractions of 0.153035 (Na), 0.845603 (I) and 0.001362 (Tl); housing, reflector and photomultiplier tube are not modelled. Its light output is equal to the energy deposited by electrons. The parameters $\delta_1$ and $\delta_2$ of Eq.~\eqref{eq:resolution} were obtained by fitting the FWHM measured by~\cite{DUCTAM201775}, while $\delta_3$ was set to 3~keV to account for electronic noise. The resulting resolution at 662~keV is 6.5\% FWHM.
\end{itemize}

\begin{table*}[!t]
\centering
\small
\caption{Detector models and pyMPPost parameters. Resolution parameters refer to Eq.~\eqref{eq:resolution}.}
\label{tab:detectors}
\begin{tabular*}{\textwidth}{@{\extracolsep{\fill}}lllllll@{}}
\toprule
Detector & Size & Density & Light output & $\delta_1$ & $\delta_2$ & $\delta_3$ \\
 & (cm) & (g\,cm$^{-3}$) & & & (MeVee$^{1/2}$) & (MeVee) \\
\midrule
trans-stilbene & $\varnothing 5.08 \times 5.08$ & 1.16 & Birks$^{a}$ & 0.078 & 0.017 & 0.028 \\
$^{4}$He (190 bar) & $\varnothing 4.95 \times 20$ & 0.0269 & linear & 0.187 & 0.2855 & 0.1175 \\
NaI(Tl) & $\varnothing 7.62 \times 7.62$ & 3.67 & linear & 0.0186 & 0.0505 & 0.003 \\
\bottomrule
\end{tabular*}
\par\smallskip
\raggedright\footnotesize $^{a}$ Eq.~\eqref{eq:birks} with $S_0 = 2.027$ and $k_B = 27.83$~mg\,cm$^{-2}$\,MeV$^{-1}$.
\end{table*}

The detectors are surrounded by void (air at negligible density). Each source is a point source located 1~cm in front of the flat face of the detector, emitting a monodirectional beam along the detector axis. The four sources were chosen to cover complementary conditions. The $^{252}$Cf spontaneous fission spectrum, described by a Watt distribution with $a = 1.025$~MeV and $b = 2.926$~MeV$^{-1}$~\cite{Pelowitz2011, Caldeira1985}, tests the reconstruction over a broad range of neutron energies, including energies above the threshold for inelastic scattering on carbon. The deuterium-deuterium (DD) fusion source, monoenergetic neutron emission at 2.45~MeV, produces a sharp edge at the maximum recoil energy, which is sensitive to any error in the reconstructed kinematics. The $^{137}$Cs source emits monoenergetic gamma rays of 661.657~keV \cite{LEBLOND2024111191} and produces the well-separated photopeak, Compton continuum and X-ray escape features that test the photon energy balance. The $^{152}$Eu source tests the photon reconstruction over a wide range of energies: it includes the 41 gamma-ray lines with emission probability of at least 0.1 per 100 decays and the K X-rays of Sm and Gd, for a total of 48 lines and 2.317 photons per decay, with energies and emission probabilities from the DDEP evaluation~\cite{Vanin2004}. Approximately 22\% of the emitted photons have energies above the pair production threshold, so that this source also exercises the reconstruction of pair production events. In each history a single photon is emitted, sampled according to the emission probabilities; correlations between photons emitted in the same decay cascade are therefore not modeled, and true-coincidence summing is not reproduced. Since the same source definition is used in both codes, this simplification does not affect the comparison.

\begin{table}[!t]
\centering
\small
\caption{Verification cases. $N$ is the number of source particles simulated with each code; the range is the light-output interval of the histograms (50 bins each).}
\label{tab:cases}
\begin{tabular*}{\columnwidth}{@{\extracolsep{\fill}}lllll@{}}
\toprule
Particle & Detector & Source & $N$ & Range (MeVee) \\
\midrule
Neutrons & trans-stilbene & $^{252}$Cf & $4\times10^{5}$ & 0.05--5 \\
Neutrons & trans-stilbene & DD & $4\times10^{5}$ & 0.05--0.7 \\
Neutrons & $^{4}$He & $^{252}$Cf & $4\times10^{5}$ & 0.05--5 \\
Neutrons & $^{4}$He & DD & $4\times10^{5}$ & 0.05--2.2 \\
Photons & NaI(Tl) & $^{137}$Cs & $1\times10^{5}$ & 0.05--0.8 \\
Photons & NaI(Tl) & $^{152}$Eu & $1\times10^{5}$ & 0.05--1.6 \\
\bottomrule
\end{tabular*}
\end{table}

\subsection{Simulation settings}
\label{sec:settings}

The same geometry, materials and sources were used in OpenMC and in MCNPX-PoliMi v2.0, which is distributed as a patch to MCNPX 2.7.0~\cite{Pelowitz2011,PoliMiManual2012}. Neutron cross sections were taken in both codes from ENDF/B-VII.1~\cite{Chadwick2011}: MCNPX-PoliMi used the \texttt{71c} ACE files for neutrons and the \texttt{04p} for photons, and OpenMC the same ACE files converted to its HDF5 format. Both codes used the TTB approximation (Section~\ref{sec:photon-reco}) and a photon energy cutoff of 1~keV. Both codes were run in analog mode, without variance reduction, as required by MCNPX-PoliMi~\cite{PoliMiManual2012}, so that all particles carry unit weight.

In MCNPX-PoliMi, all sources were defined through the standard SDEF card (IPOL(1)~=~0). In particular, the $^{252}$Cf source was described by the same Watt spectrum used in OpenMC rather than by the built-in spontaneous fission source of MCNPX-PoliMi, so that both codes emit a single neutron per history and no fission multiplicity is sampled. The collision file was written for a single detector cell (IPOL(7)~=~1, IPOL(8) equal to the cell number). MCNPX-PoliMi writes to the collision file only the energy depositions that exceed the thresholds RPOL(1) for neutrons and RPOL(2) for photons~\cite{PoliMiManual2012}. In the neutron cases these were set to 0.01~MeV and 100~MeV, so that neutron collisions depositing less than 10~keV and all photon collisions are not written. \texttt{TrackAnalyzer} applied the same thresholds (Section~\ref{sec:collision-file}): 0.01~MeV on the deposited energy in the neutron cases and none in the photon cases, and it wrote only the collisions in the detector cell. In the photon cases both thresholds were set to zero and no collision was discarded. In OpenMC, photon transport was enabled also in the neutron simulations, but only neutron tracks were analysed.

Each MCNPX-PoliMi case was run as 100 independent simulations of $4\times10^{3}$ (neutrons) or $1\times10^{3}$ (photons) histories, with a different random number seed for each run, for a total equal to the number of histories simulated with OpenMC. pyMPPost was run on each MCNPX-PoliMi collision file, and the pulses of all files were merged and normalized to the total number of histories, including files without detector events. Only the interactions of the analysed particle were passed to pyMPPost, so that pulses mixing neutron and photon contributions are avoided and the two codes are compared on the same basis.

For each case, the light-output distributions obtained with the two codes were histogrammed with 50 equal-width bins over the ranges listed in Table~\ref{tab:cases} and normalized per source particle. The content of bin $j$ is $y_j = N_j/N_{\mathrm{s}}$, where $N_j$ is the number of pulses in the bin and $N_{\mathrm{s}}$ the number of source particles, and its statistical uncertainty is $\sigma_j = \sqrt{N_j}/N_{\mathrm{s}}$. The two distributions are compared bin by bin through the ratio
\begin{equation}
R_j = \frac{y_j^{\mathrm{OpenMC}}}{y_j^{\mathrm{PoliMi}}},
\qquad
\sigma_{R_j} = R_j\sqrt{\left(\frac{\sigma_j^{\mathrm{OpenMC}}}{y_j^{\mathrm{OpenMC}}}\right)^{2} + \left(\frac{\sigma_j^{\mathrm{PoliMi}}}{y_j^{\mathrm{PoliMi}}}\right)^{2}},
\label{eq:ratio}
\end{equation}
where the uncertainties of the two codes are combined in quadrature, since the two simulations are statistically independent. The ratio is computed only for bins with non-zero content in both codes. Agreement between the codes corresponds to ratios compatible with unity within their uncertainties.

\section{Results}
\label{sec:results}

\begin{figure*}[!t]
\centering
\includegraphics[width=0.49\textwidth]{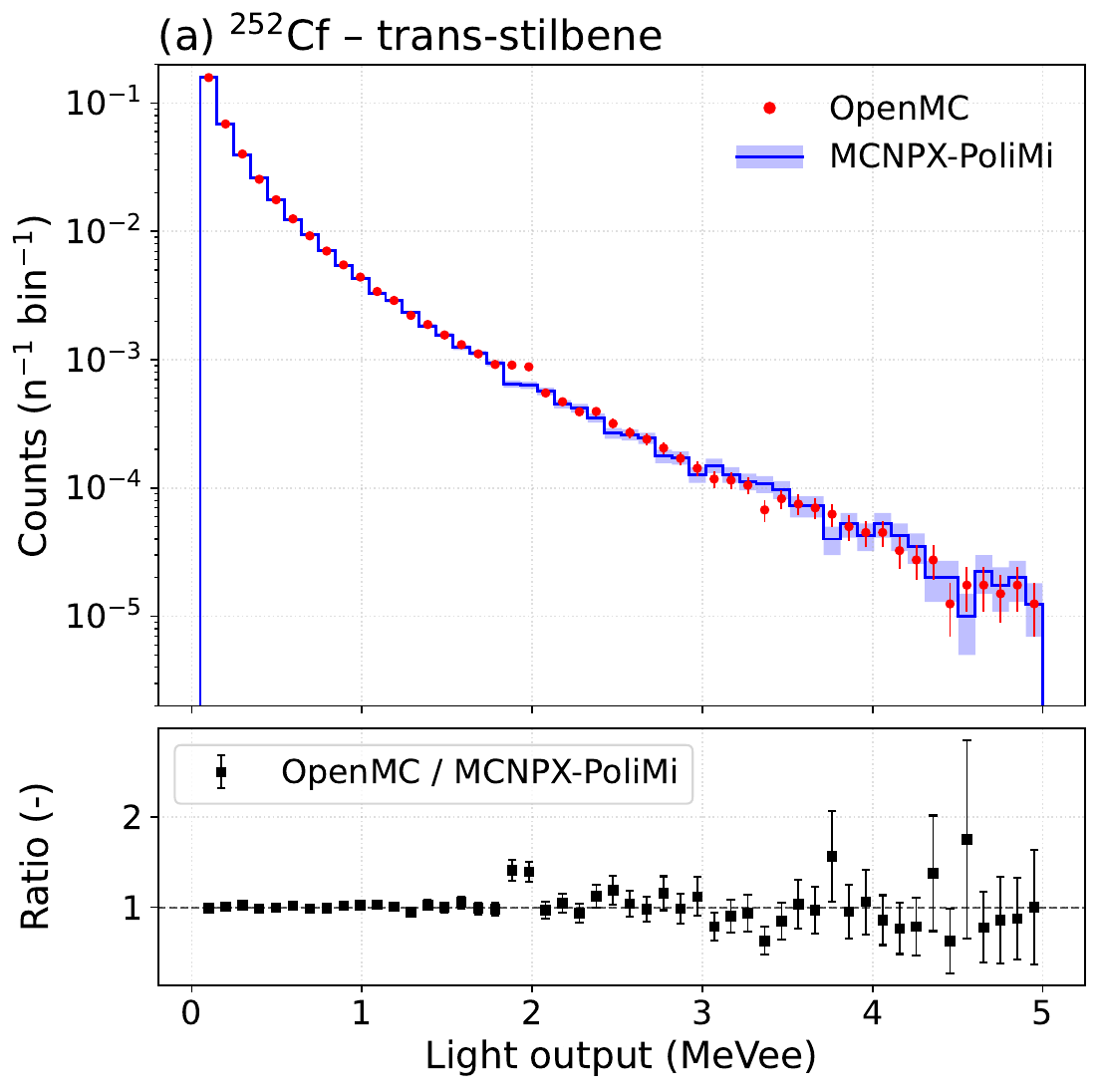}\hfill
\includegraphics[width=0.49\textwidth]{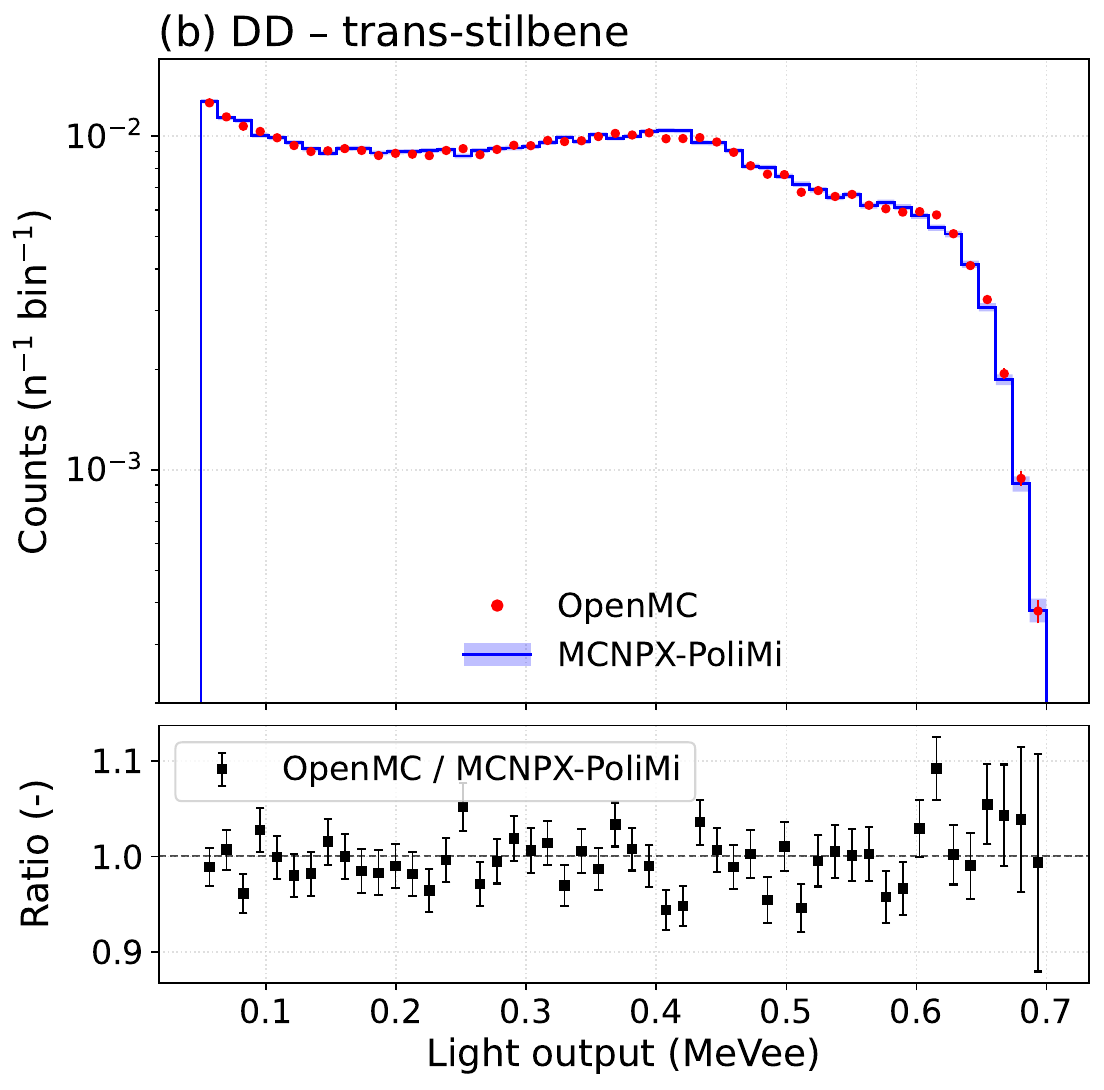}\\[2mm]
\includegraphics[width=0.49\textwidth]{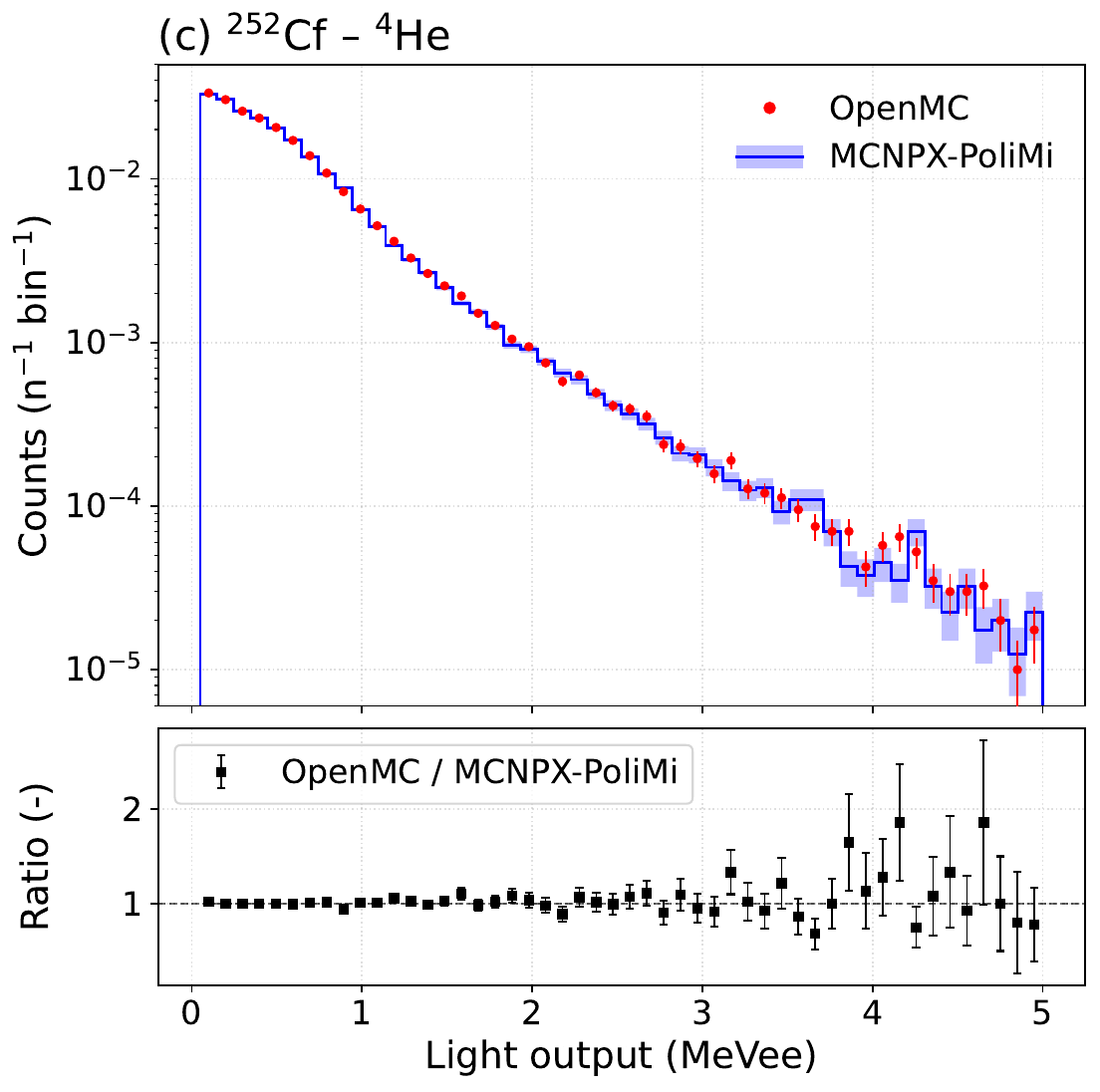}\hfill
\includegraphics[width=0.49\textwidth]{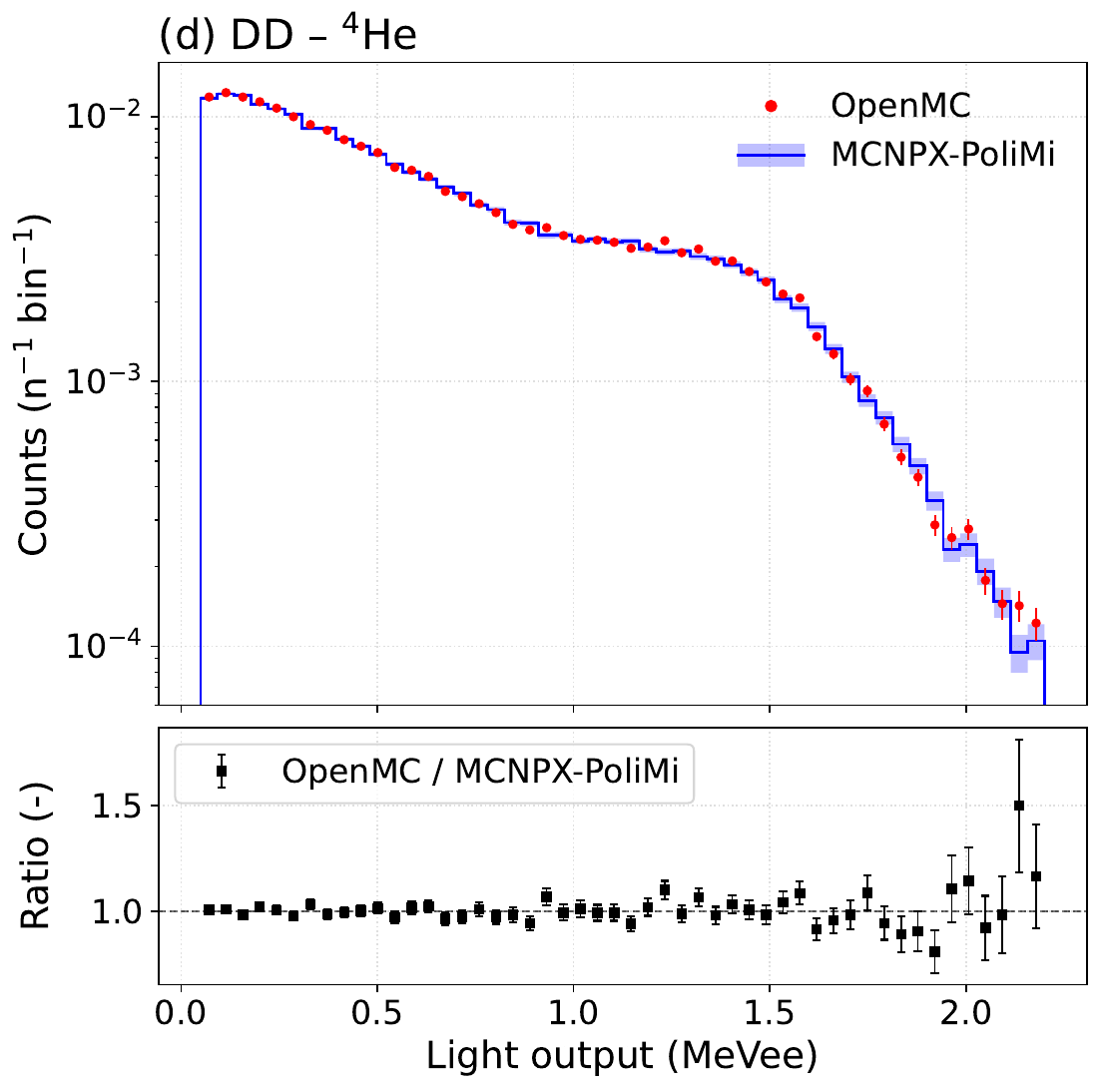}
\caption{Light-output distributions for the neutron cases: (a) $^{252}$Cf and (b) DD source on the trans-stilbene detector; (c) $^{252}$Cf and (d) DD source on the $^{4}$He detector. Upper plots: MCNPX-PoliMi (histogram with $\pm1\sigma$ band) and OpenMC (points with $\pm1\sigma$ error bars), normalized per source neutron. Lower plots: ratio between OpenMC and MCNPX-PoliMi with its uncertainty.}
\label{fig:neutron}
\end{figure*}

\begin{figure*}[!t]
\centering
\includegraphics[width=0.49\textwidth]{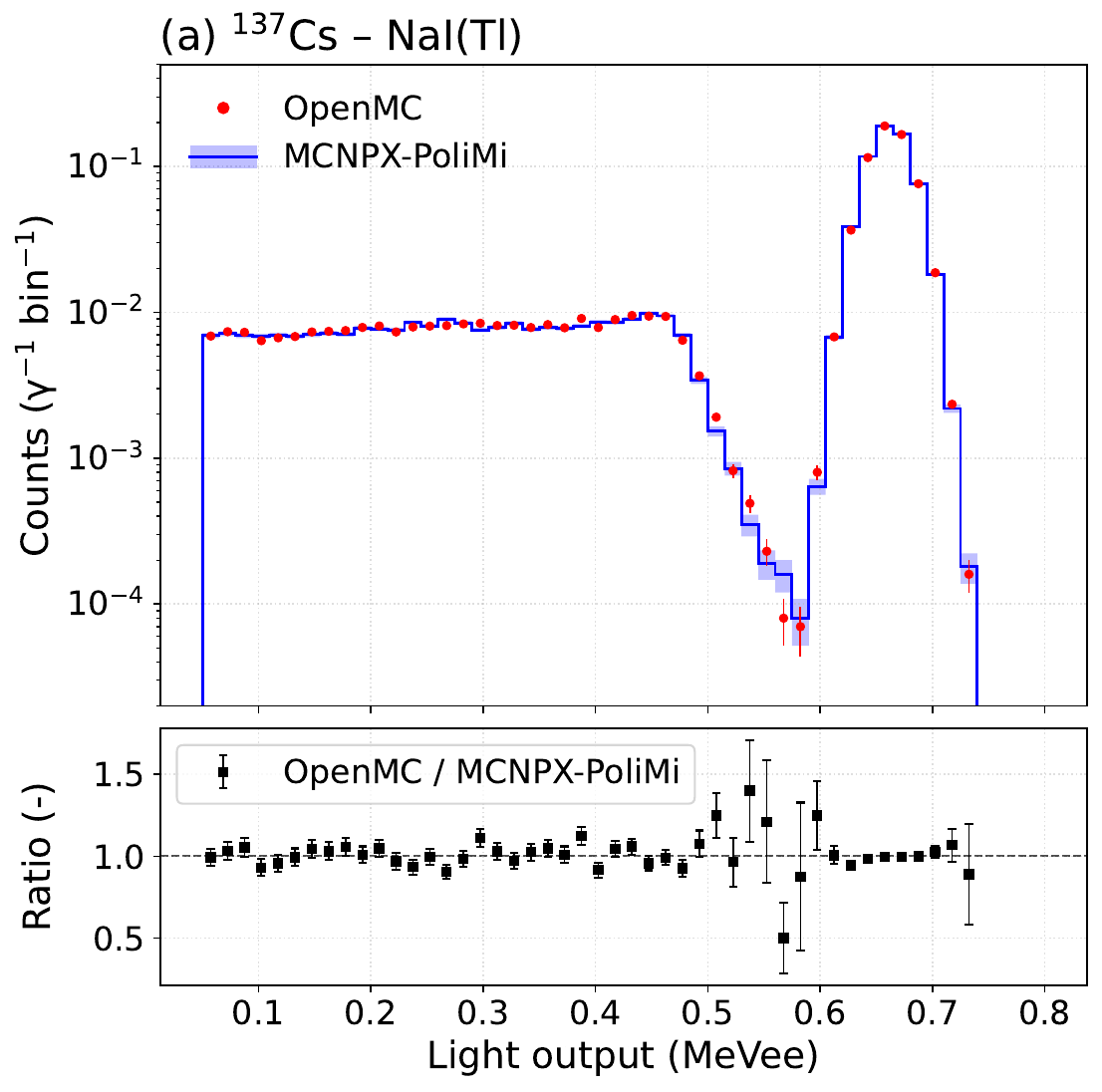}\hfill
\includegraphics[width=0.49\textwidth]{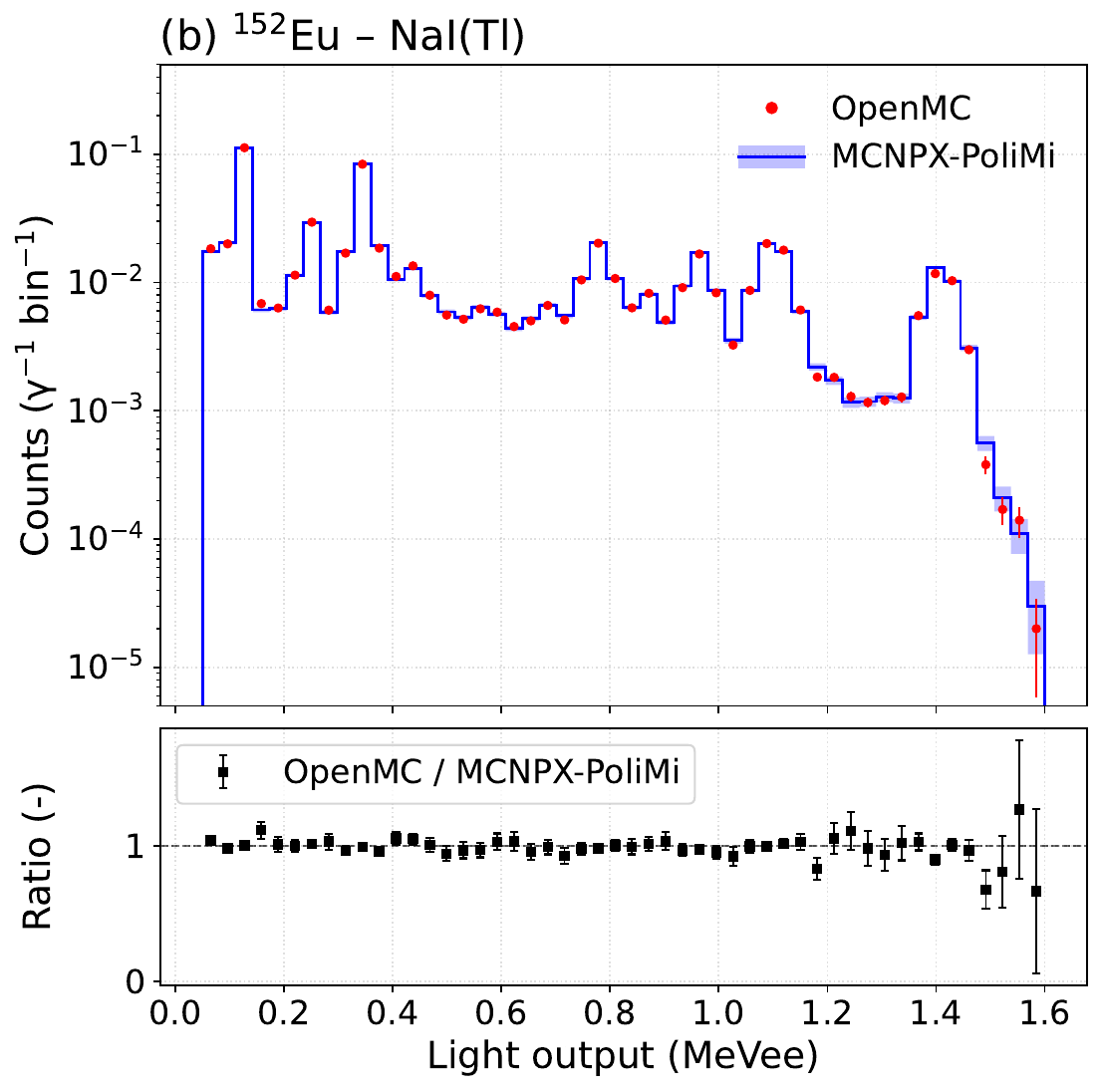}
\caption{Light-output distributions for the photon cases on the NaI(Tl) detector: (a) $^{137}$Cs and (b) $^{152}$Eu source. Upper plots: MCNPX-PoliMi (histogram with $\pm1\sigma$ band) and OpenMC (points with $\pm1\sigma$ error bars), normalized per source photon. Lower plots: ratio between OpenMC and MCNPX-PoliMi with its uncertainty.}
\label{fig:photon}
\end{figure*}

The light-output distributions obtained with OpenMC and \texttt{TrackAnalyzer} are compared with those obtained with MCNPX-PoliMi in Fig.~\ref{fig:neutron} for the four neutron cases and in Fig.~\ref{fig:photon} for the two photon cases. Since the collision files of both codes are processed by pyMPPost with identical parameters, any difference between the distributions originates either in the transport or in the reconstruction of the interactions from the OpenMC tracks.

\subsection{Neutron cases}
\label{sec:results-neutron}

In all four neutron cases, the ratio between OpenMC and MCNPX-PoliMi is compatible with unity over most of the light-output range, with deviations of a few percent in the bins where the statistical uncertainty is of the same order. For trans-stilbene (Fig.~\ref{fig:neutron}a,b), this holds up to about 1.8~MeVee with the $^{252}$Cf source and over the whole distribution with the DD source, including the edge corresponding to the full neutron energy of 2.45~MeV. For $^{4}$He (Fig.~\ref{fig:neutron}c,d), the agreement extends up to about 3~MeVee with the $^{252}$Cf source and up to 1.7~MeVee with the DD source. At higher light output the counts decrease by several orders of magnitude, and the ratio fluctuates around unity within uncertainties of 20--100\%.

Only one deviation exceeds the statistical uncertainties. With the $^{252}$Cf source on trans-stilbene, OpenMC exceeds MCNPX-PoliMi by about 40\% in the two bins between 1.83 and 2.03~MeVee (about four standard deviations). Part of this excess is due to inelastic collisions on carbon reconstructed as elastic scattering on hydrogen (Section~\ref{sec:discussion} and Appendix~\ref{app:collision-track}). In the other neutron cases the few bins outside two standard deviations are isolated, and $\chi^2/\mathrm{ndf}$ ranges from 0.82 to 1.25, consistent with statistical fluctuations.

\subsection{Photon cases}
\label{sec:results-photon}

For the NaI(Tl) detector (Fig.~\ref{fig:photon}), the two codes agree within 1\% in the $^{137}$Cs photopeak, which contains most of the counts, within the statistical uncertainty of 1--4\% in the $^{152}$Eu photopeaks, and within 5--10\% in the Compton continua. This applies to the $^{137}$Cs spectrum, with its Compton edge at 0.48~MeVee and photopeak at 0.662~MeVee, and to the multi-line $^{152}$Eu spectrum, from the K X-ray peak at about 40~keV up to the photopeak of the 1.408~MeV line. Larger deviations occur only in bins with very few counts, such as the valley between the Compton edge and the photopeak of $^{137}$Cs and the region above 1.45~MeVee for $^{152}$Eu, and are compatible with unity within their uncertainties.

\subsection{Integral comparison}
\label{sec:results-integral}

Table~\ref{tab:integral} compares the total number of pulses per source particle in the histogram range. The two codes agree within 0.6\% in all cases, and all the ratios are compatible with unity within 1.4 standard deviations

\begin{table}[!t]
\centering
\footnotesize
\setlength{\tabcolsep}{4pt}
\caption{Total number of pulses per source particle in the histogram range of Table~\ref{tab:cases}, and ratio between OpenMC and MCNPX-PoliMi. Uncertainties are statistical (one standard deviation).}
\label{tab:integral}
\begin{tabular*}{\columnwidth}{@{}l@{\extracolsep{\fill}}ccc@{}}
\toprule
Case & OpenMC & MCNPX-PoliMi & Ratio \\
\midrule
Stilbene, $^{252}$Cf & $0.3697 \pm 0.0010$ & $0.3700 \pm 0.0010$ & $0.999 \pm 0.004$ \\
Stilbene, DD & $0.4042 \pm 0.0010$ & $0.4061 \pm 0.0010$ & $0.995 \pm 0.003$ \\
$^{4}$He, $^{252}$Cf & $0.2205 \pm 0.0007$ & $0.2193 \pm 0.0007$ & $1.005 \pm 0.005$ \\
$^{4}$He, DD & $0.2113 \pm 0.0007$ & $0.2109 \pm 0.0007$ & $1.002 \pm 0.005$ \\
NaI(Tl), $^{137}$Cs & $0.8466 \pm 0.0029$ & $0.8499 \pm 0.0029$ & $0.996 \pm 0.005$ \\
NaI(Tl), $^{152}$Eu & $0.5970 \pm 0.0024$ & $0.5985 \pm 0.0024$ & $0.998 \pm 0.006$ \\
\bottomrule
\end{tabular*}
\end{table}
\section{Discussion}
\label{sec:discussion}

Since both codes used the same ENDF/B-VII.1 neutron data, the agreement shown in Section~\ref{sec:results}, within 1\% in the total number of pulses and within a few percent in the light-output distributions wherever statistics allow, reflects the combined accuracy of OpenMC transport and of the reconstruction from the track file. The reference is itself supported by measurements: for the trans-stilbene detector, MCNPX-PoliMi with MPPost and the same light-output function reproduced the measured response to a $^{252}$Cf source within 3\%~\cite{SHIN201936}. Within the scope of the cases considered, OpenMC coupled with \texttt{TrackAnalyzer} therefore provides an alternative to MCNPX-PoliMi for the simulation of detector responses with pyMPPost. Since the comparison involves the whole chain, from transport to pulse formation, the same models and input files can also serve as a reference for verifying future versions of OpenMC or of the coupler.

One residual difference deserves comment. The excess of OpenMC near 2~MeVee in the $^{252}$Cf--trans-stilbene case is due, for up to about half of its magnitude, to inelastic collisions on carbon that are compatible within the tolerance with elastic scattering on hydrogen: about $2.4\times10^{-4}$ per source neutron, converted by Eq.~\eqref{eq:birks} into 1.89--1.94~MeVee, which accounts for the position of the excess (~\ref{app:collision-track}); the remaining part could not be attributed. Since scattering on a discrete level is also a two-body process, the kinematic test of Eq.~\eqref{eq:elastic} could be extended to include the excited levels of carbon as additional candidates with the corresponding reaction Q values, so that these collisions would be recognized and rejected; this extension was not applied in this work.

The agreement also requires the two collision files to be written with consistent options. MCNPX-PoliMi does not write the collisions that deposit less than the RPOL thresholds, and these collisions, although individually small, add light to the pulses in which pyMPPost sums them. Their effect is largest where the light-output distribution varies steeply, as near the recoil edge of a monoenergetic source, and in detectors with a linear response, where the light of low-energy depositions is not reduced by quenching. For this reason \texttt{TrackAnalyzer} applies the same thresholds as the reference input (Section~\ref{sec:collision-file}), and the RPOL settings of the MCNPX-PoliMi input should be reproduced whenever the two codes are compared. When the response is compared with measurements, the threshold is not a property of the detector: it should be set to zero, or low enough for the discarded light to be negligible.

The reconstruction has some limitations that should be considered when applying it to other problems. For neutrons, only elastic scattering is written. This has little effect for the sources considered here, but it becomes relevant at higher energies, for example for 14.1~MeV neutrons from deuterium-tritium (DT) sources, where inelastic scattering and $(n,\alpha)$ reactions on carbon contribute significantly to the light output of organic scintillators. For photons, pair production was not specifically verified (Section~\ref{sec:photon-reco}). Photonuclear reactions occurring in the detector material are not identified and would be treated as photoelectric absorptions, with an incorrect deposited energy, since only the energy of secondary photons is subtracted; this has no effect on the cases considered here, whose photon energies are below the photonuclear thresholds of the detector nuclides, but it should be addressed before applying the method to high-energy photon beams.

Other approximations, described in Section~\ref{sec:methodology}, are common to both codes or concern the detector models rather than the coupling: the TTB treatment of electrons, which may affect high-energy photons in small detectors, the single-photon $^{152}$Eu source, the bare detector models, the extrapolation of the trans-stilbene light output above 4.57~MeV, and the anisotropy of crystalline trans-stilbene~\cite{SCHUSTER201795}, which pyMPPost does not model. Similarly, the light output of electrons is assumed to be proportional to the deposited energy, whereas NaI(Tl) and other scintillators exhibit a non-proportional response~\cite{Payne2011,Breitenmoser2023}. They do not affect the code-to-code comparison, but must be revisited when comparing with measurements.

The computational cost of the coupling is small compared with that of the transport. For the $^{252}$Cf--trans-stilbene case ($4\times10^{5}$ histories, Apple M2 processor), the OpenMC simulation took 352~s with 8 threads, about 0.8 CPU-hours; the conversion of the track file with \texttt{TrackAnalyzer}, which runs as a single process, took 140~s of CPU time (0.04 CPU-hours, 0.35~ms per history) with a peak memory of 0.46~GB; and pyMPPost took about 6~s. The main practical constraint is the size of the track file, which was 497~MB (1.24~kB per history) and grows linearly with the number of histories; as noted in Ref.~\cite{Saliba2026}, storing all tracks can require hundreds of gigabytes in routine cases. The collision file is about four times smaller (114~MB). For large simulations, the run can be split into independent batches with different random seeds, each converted into a collision file immediately after the transport and then deleted. The size of the track file also depends on the secondary particles that are tracked: in the neutron cases, photon transport was enabled and the photon tracks, which are not used by the analysis, were also written.

The coupling is complementary to the collision tracking feature of the official OpenMC release~\cite{Saliba2026}: where it is available, the same writer of the MCNPX-PoliMi format can take its output as input, and a function for this purpose is included in the repository of this work. Finally, the verification presented here concerns the pulse-height module of pyMPPost. The collision file also contains the time and the cell of each interaction, which are required by the cross-correlation and multiplicity modules, but these modules were not verified in this work. Beyond nonproliferation and safeguards, the coupling can be applied wherever the response of a scintillation detector has to be predicted with an open-source code, for example in the design of radiation monitoring systems for research reactors, accelerators and fusion facilities, and in teaching radiation detection.
\section{Conclusions}\label{sec:conclusion}

This work presented a coupling between the open-source Monte Carlo code OpenMC and the detector response post-processor pyMPPost. The coupler, \texttt{TrackAnalyzer}, reconstructs the interactions in the detector from the OpenMC particle tracks and writes them in the MCNPX-PoliMi collision file format, so that the detector models of pyMPPost can be used without modification of either code. The target nuclide of neutron collisions, which the track file does not store, is reconstructed from two-body elastic kinematics; this makes the method applicable to detectors with several nuclides and a nonlinear light response, and to any OpenMC version that writes particle tracks, including development branches where the collision tracking feature is not available.

The code-to-code verification against MCNPX-PoliMi, with identical pyMPPost parameters, covered neutrons and photons on three detectors with different characteristics: trans-stilbene, $^{4}$He and NaI(Tl). The total number of pulses agreed within 0.6\% in all cases, and the light-output distributions agreed within the statistical uncertainty wherever statistics allow, provided that both collision files are written with the same threshold on the deposited energy. The only residual difference, an excess near 2~MeVee in the $^{252}$Cf--trans-stilbene case, is explained in part by inelastic collisions on carbon reconstructed as elastic scattering on hydrogen. The additional computational cost of the reconstruction is small compared with that of the transport, and the main practical constraint is the size of the track file.

Several developments follow from this work. The kinematic test can be extended to the excited levels of carbon, so that inelastic collisions are rejected or written correctly, which is necessary for higher-energy sources such as DT neutrons. The cross-correlation and multiplicity modules of pyMPPost, which use the time and cell information already written in the collision file, remain to be verified. Where the collision tracking feature of the official release is available, its output can be passed directly to the same writer of the MCNPX-PoliMi format. Finally, the coupling provides the detector response needed to validate the photonuclear physics of OpenMC against measurements, which is the next step of this work.

\section*{CRediT authorship contribution statement}
\textbf{Lorenzo Loi:} Conceptualization, Software, Writing -- original draft.
\textbf{David Breitenmoser:} Methodology, Writing -- original draft.
\textbf{Shaun D. Clarke:} Methodology, Supervision, Writing -- review \& editing.
\textbf{Sara A. Pozzi:} Methodology, Supervision, Writing -- review \& editing.
\textbf{Antonio Cammi:} Supervision, Writing -- review \& editing.

\section*{Data availability}
The source code of \texttt{TrackAnalyzer}, together with the OpenMC models, the pyMPPost input files and the processed results used to produce the figures and tables of this article, is openly available at \url{https://github.com/ERMETE-Lab/openmc-to-pymppost} under the MIT license.

\section*{Acknowledgements}
This research was supported in part through computational resources and services provided by Advanced Research Computing at the University of Michigan, Ann Arbor.

\section*{Declaration of competing interest}
The authors declare that they have no known competing financial interests or personal relationships that could have appeared to influence the work reported in this paper.

\section*{Funding}
This research did not receive any specific grant from funding agencies in the public, commercial, or not-for-profit sectors.

\newpage
\appendix

\section{Event-by-event verification of the neutron collision reconstruction}
\label{app:collision-track}

Starting from version 0.15.3, OpenMC can record individual collisions through the collision tracking feature \cite{Saliba2026}. For every collision it stores the reaction type (ENDF MT number), the target nuclide, the energy of the particle before the collision and the energy it loses. It provides a direct reference for the kinematic reconstruction of Section~\ref{sec:neutron-reco}: when both particle tracks and collision tracking are enabled in the same simulation, every collision is described twice, once by the reconstruction and once by the transport code itself.

\subsection{Setup}

The $^{252}$Cf--trans-stilbene case of Section~\ref{sec:cases} was simulated with the official release of OpenMC 0.15.3 (commit \texttt{27e38e89}) and $10^{5}$ source neutrons, with the same geometry, material composition and source as in the main analysis. Photon transport was disabled. Cross sections were taken from the ENDF/B-VII.1 library distributed in HDF5 format for OpenMC \cite{Chadwick2011}, i.e.\ the same evaluation used in the main simulations but processed independently. Collision tracking was restricted to the detector material, without any filter on the reaction type, so that elastic, inelastic and absorption events were all recorded.

The tracks were processed with the reconstruction settings used for the results of Section~\ref{sec:results} (incident-energy threshold 10~eV, residual tolerance 0.05, candidate targets $^{1}$H and C), but without the threshold on the deposited energy, so that every reconstructed collision could be compared with the recorded one. The reconstructed collisions were matched to the recorded ones by source particle number and collision position, rounded to $10^{-7}$~cm. No ambiguous matches occurred.

\subsection{Results}

Table~\ref{tab:ct-counts} compares the true reaction of each collision with the outcome of the reconstruction, and Table~\ref{tab:ct-summary} summarizes the corresponding figures of merit.

\begin{table*}[!t]
\centering
\scriptsize
\caption{Collisions in the trans-stilbene detector ($10^{5}$ $^{252}$Cf source neutrons): true reaction from OpenMC collision tracking (rows) versus outcome of the track-based reconstruction (columns). Rejection reasons: incident energy below the 10~eV threshold, thermal up-scattering, near-forward scattering ($E'/E > 1-10^{-4}$), no target compatible with elastic kinematics, best residual above tolerance.}
\label{tab:ct-counts}
\begin{tabular*}{\textwidth}{@{\extracolsep{\fill}}lrrrrrrrrr@{}}
\toprule
& & & \multicolumn{5}{c}{Rejected} & \multicolumn{2}{c}{Written as} \\
\cmidrule(lr){4-8}\cmidrule(l){9-10}
True reaction & Total & Not det. & $E<$\,thr. & Up-sc. & Fwd. & No target & Resid. & $^{1}$H & C \\
\midrule
$^{1}$H elastic & 144265 & 0 & 1833 & 1340 & 17 & 591 & 3112 & 137360 & 12 \\
C elastic & 76411 & 0 & 702 & 339 & 43 & 146 & 370 & 0 & 74811 \\
C inelastic, discrete & 396 & 0 & 0 & 0 & 0 & 159 & 215 & 22 & 0 \\
C inelastic, continuum & 7 & 0 & 0 & 0 & 0 & 2 & 3 & 2 & 0 \\
Absorption ($^{1}$H and C) & 93 & 93 & 0 & 0 & 0 & 0 & 0 & 0 & 0 \\
\midrule
Total & 221172 & 93 & 2535 & 1679 & 60 & 898 & 3700 & 137384 & 74823 \\
\bottomrule
\end{tabular*}
\end{table*}

\begin{table}[!t]
\centering
\small
\caption{Figures of merit of the reconstruction, from Table~\ref{tab:ct-counts}. Energy refers to the energy lost by the neutron in the collisions considered.}
\label{tab:ct-summary}
\begin{tabular*}{\columnwidth}{@{\extracolsep{\fill}}lrr@{}}
\toprule
Collisions & Events & Energy \\
\midrule
$^{1}$H elastic, correct nuclide & 95.2\% & 99.94\% \\
C elastic, correct nuclide & 97.9\% & $>$99.99\% \\
Elastic, wrong nuclide & 12 & $<$0.01\% \\
Inelastic, rejected & 94.0\% & -- \\
Inelastic, written as $^{1}$H & 24 & 128.8~MeV \\
Written, correctly identified & 99.98\% & 99.82\% \\
\bottomrule
\end{tabular*}
\end{table}

All scattering collisions in the detector are detected from the direction change, and no collision is detected where none occurred. The 93 collisions without a direction change are all absorptions, which are not written by construction. For every written event, the position, time and deposited energy coincide exactly with the values recorded by OpenMC, which confirms the reading of the track states described in Section~\ref{sec:tracks}.

Elastic collisions are identified correctly in 95.2\% of the cases for $^{1}$H and 97.9\% for carbon, and only 12 of them (all with negligible energy loss) are assigned to the wrong nuclide. The remaining elastic collisions are rejected. Most of them occur at low incident energy: below the 10~eV threshold, with thermal up-scattering, or at energies of tens of eV, where the thermal motion of the target is not negligible and the kinematics of a target at rest no longer applies. As a consequence, the rejected events carry only 0.06\% of the energy lost in elastic collisions on $^{1}$H, and 99.95\% of the energy lost in elastic collisions is written and attributed to the correct nuclide. Since the light output of the detector is driven by the deposited energy, the losses in number of events have no practical effect on the simulated light-output distribution.

Inelastic collisions on carbon are rejected in 94.0\% of the cases, mostly because no elastic target is compatible with the observed kinematics. However, 24 inelastic collisions (21 of them on the first excited level of $^{12}$C at 4.44~MeV) fall within the residual tolerance of $^{1}$H elastic scattering and are written as proton recoils. Their recorded energy loss, about 4.8~MeV for the first level, includes the excitation energy of the target nucleus, which in a real detector would be carried away by a de-excitation photon. These events are few, corresponding to 0.18\% of the written energy and $2.4\times10^{-4}$ events per source neutron, but pyMPPost converts them into proton light concentrated in a narrow region: with the light-output function of Eq.~\eqref{eq:birks}, a 4.82--4.88~MeV proton produces 1.89--1.94~MeVee. These values fall in the two bins where OpenMC exceeds MCNPX-PoliMi in the $^{252}$Cf--trans-stilbene comparison (Fig.~\ref{fig:neutron}a).

This effect is limited to neutrons above the inelastic threshold of carbon (about 4.8~MeV) and therefore concerns only the $^{252}$Cf--trans-stilbene case among those considered. With the 2.45~MeV DD source no inelastic channel is open in trans-stilbene, and in $^{4}$He elastic scattering is the only open channel below 20~MeV. The script used for this comparison is available in the repository of this work.

\bibliography{bibliography}

\end{document}